\documentclass[aps, prd, reprint, longbibliography, nofootinbib,superscriptaddress, floatfix]{revtex4-2}
\pdfoutput=1

\usepackage{amsmath,amssymb,amsfonts}
\usepackage{mathrsfs}
\usepackage{bbm}
\usepackage{slashed}
\usepackage{graphicx}
\usepackage{comment}
\usepackage{verbatim}
\usepackage{balance}
\usepackage[T1]{fontenc}
\usepackage[utf8]{inputenc}
\usepackage[colorlinks]{hyperref}
\usepackage{mathbbol}
\usepackage[dvipsnames]{xcolor}
\usepackage[normalem]{ulem}
\usepackage{svg}
\usepackage{flushend}
\usepackage{balance}
\usepackage{physics}
\usepackage{graphicx}
\usepackage{dcolumn}
\usepackage{bm}
\usepackage{tensor}
\usepackage{comment}
\usepackage[utf8]{inputenc} 
\usepackage{graphicx,color,overpic,mathtools}
\usepackage{amsthm,amsmath,amssymb,hyperref,mathrsfs}
\usepackage{braket,bm,bbm}
\usepackage{cancel}
\PassOptionsToPackage{normalem}{ulem}
\usepackage{ulem} 
\usepackage{physics}
\usepackage{float}
\usepackage[english]{babel}
\usepackage{graphicx}
\hypersetup{
    colorlinks=false,
    pdfborder={0 0 0},
}

\newcommand{\F}{\mathcal{F}}

\begin{document}

\title{Diverging black hole entropy from quantum infrared non-localities}

\author{Alessia Platania}
\email[]{aplatania@perimeterinstitute.ca}
\affiliation{
Perimeter Institute for Theoretical Physics, 31 Caroline Street North, Waterloo, ON N2L 2Y5, Canada
}
\affiliation{Nordita, KTH Royal Institute of Technology and Stockholm University, Hannes Alfv\'ens v\"ag 12, SE-106 91 Stockholm, Sweden}

\author{Jaime Redondo--Yuste}
\email[]{jaime.redondo.yuste@nbi.ku.dk}
\affiliation{
Perimeter Institute for Theoretical Physics, 31 Caroline Street North, Waterloo, ON N2L 2Y5, Canada
}
\affiliation{Niels Bohr International Academy, Niels Bohr Institute, Blegdamsvej 17, 2100 Copenhagen, Denmark}

\begin{abstract}
    Local higher-derivative corrections to the Einstein-Hilbert action yield sub-leading corrections to the Bekenstein-Hawking area law. Here we show that if the quantum effective action comprises a certain class of infrared non-localities, the entropy of large black holes generally diverges to either positive or negative infinity. In such theories, large spherically symmetric black holes would be either highly chaotic or thermodynamically impossible, respectively. In turn, this puts strong constraints on the Laurent expansion of the form factors in the effective action.
\end{abstract}

\maketitle



The theoretical prediction that black holes emit thermal radiation~\cite{Hawking:1974rv,Hawking:1975vcx} is considered one of Stephen Hawking's most important contributions to our  understanding of black holes. Hawking radiation has two crucial consequences: \emph{(i)} it entails that black holes evaporate, as the emission causes a reduction in their mass and rotational energy, and \emph{(ii)} the entropy of a large black hole is, to leading order, proportional to its area~\cite{Bekenstein:1972tm,Hawking:1975vcx}
\begin{equation}\label{eq:area-law}
    S_{BH}=\frac{\mathcal{A}}{4l_P^2} \,,
\end{equation}
with $l_P$ being the Planck length. Corrections to this classical formula parametrize deviations from General Relativity (GR).

The first law of black hole mechanics connects the perturbations of the black hole mass, its angular momentum, and a combination of the surface gravity and a geometrical quantity that is later given the meaning of an entropy. The original derivation~\cite{Bardeen:1973gs} used very explicit properties of the Einstein field equations, so a natural question was whether such a statement could be extended to any theory of gravity. Wald's seminal work~\cite{Wald:1993nt} demonstrated that the first law holds for generic gravitational actions~$S$, provided that some minimal assumptions on the (classical or effective) Lagrangian, such as diffeomorphism invariance, are fulfilled.

On the theoretical side, the form of the effective action depends on quantum gravitational effects, and potentially on the specific ultraviolet completion of gravity. In particular, owed to the quantum nature of the effective action, a common feature is the presence of non-localities. The latter could be exponential functions of the d'Alembert operator, as in non-local gravity~\cite{Modesto:2017sdr}, hyperbolic tangents, as in some asymptotically safe models~\cite{Draper:2020bop}, operators of the Polyakov type, as in string theory~\cite{Polyakov:1987zb,Aharony:2011gb} and in the high-energy limit of loop quantum gravity~\cite{Borissova:2022clg}, or even higher inverse powers of the d'Alembert operator, as it could happen in causal dynamical triangulations~\cite{Knorr:2018kog}. Such non-localities also play a role on the phenomenological side. Specifically, inverse powers of the d'Alembert operator seem to be important in cosmological settings~\cite{Wetterich:1997bz, Deser:2007jk, Nersisyan:2016jta,Vardanyan:2017kal,Belgacem:2020pdz}, as an alternative explanation to dark energy, as well as in black hole physics~\cite{Knorr:2022kqp}, to accommodate for regular black holes stemming from a principle of least action in quantum gravity. 

The entropy of black holes in the presence of local higher-derivative terms has been widely studied~\cite{Conroy:2015wfa,Conroy:2015nva,Giaccari:2015vfh,Myung:2017axf,Calmet:2021lny}. As expected, adding such terms to the Einstein-Hilbert action leads to sub-leading corrections to the area law~\eqref{eq:area-law}. On dimensional grounds, one might expect that similar corrections stemming from inverse powers of the d'Alembert operator would dominate over the Bekenstein-Hawking scaling. The main motivation of our Letter is to test this expectation. 

We provide analytical and numerical evidence that if the effective action contains infrared non-localities, not only the resulting corrections are dominant, but also divergent. 
The entropy of spherically symmetric configurations is thus generically divergent, barring unlikely cancellations of the infinities associated with each non-local term in the effective action. Depending on the sign of the divergence, black holes in such theories would be characterized either by infinitely many microstates or by none. This paradoxical behavior points to novel constraints on the Laurent expansion of the form factors in the effective action.


\emph{Setup} --- We shall focus on asymptotically flat spacetimes. The gravitational effective action of a generic diffeomorphism-invariant theory involving only the metric is
\begin{equation} \label{action}
    \Gamma_{\mathrm{eff}}=\int d^4 x \sqrt{-g} \,  \left(\frac{R}{16\pi G}+\mathcal{L}_\mathrm{HD} \right), 
\end{equation}
where the Einstein-Hilbert (EH) part is complemented by the higher-derivative (HD) Lagrangian
\begin{align}
    \mathcal{L}_\mathrm{HD}&=\frac{1}{16\pi G}\left(R\F_1(\Box)R+R_{\mu\nu}\F_2(\Box)R^{\mu\nu}\right. \label{hd-action}\\
    & \left. \qquad +R_{\mu\nu\rho\sigma}\F_3(\Box)R^{\mu\nu\rho\sigma}+\mathcal{O}(R^3)\right) \nonumber.
\end{align}
Above, $G$ is the Newton constant and $\Box=-g^{\mu \nu} D_{\mu} D_{\nu}$ is the d'Alembert operator constructed from the Levi-Civita connection of the metric~$g$. The form factors~$\mathcal{F}_i$\footnote{One could eliminate one of the $\mathcal{F}_i$ by exploiting a generalized version of the Gauss-Bonnet identity. However, the Gauss-Bonnet term contributes to the Wald entropy even in $d=4$~\cite{Azeyanagi:2009wf}. Thus, the Gauss-Bonnet identity should not be used in this context.} result from integrating out quantum gravitational fluctuations in the functional integral, are typically non-local, and at one loop they ought to match the logarithmic behavior encountered in~\cite{Calmet:2021lny}. Their form and properties are strictly tied to the specific ultraviolet completion of gravity~\cite{Knorr:2021iwv}. 

Computing the exact entropy corrections generated by these non-local form factors is generally involved, but within the annulus of convergence, where the form factors are holomorphic, one can exploit their Laurent expansion
\begin{equation}\label{eq:Laurent}
    \mathcal{F}_i(\Box)=\sum_{n=-\infty}^\infty c_n \Box^n\,.
\end{equation}
In practical applications only a few of the negative-degree terms play a role in quantum gravity and in cosmological models~\cite{Polyakov:1987zb,Wetterich:1997bz, Deser:2007jk, Nersisyan:2016jta,Vardanyan:2017kal,Knorr:2018kog,Belgacem:2020pdz,Knorr:2022kqp,Borissova:2022clg}. In particular, such low-order terms appear to be compatible with cosmological observations~\cite{Vardanyan:2017kal,Belgacem:2020pdz}.
Entropy corrections coming from the first positive-degree terms ($n\geq0$) in this Laurent series have been computed in~\cite{Conroy:2015wfa,Conroy:2015nva,Giaccari:2015vfh,Myung:2017axf,Calmet:2021lny}. Given the relevance of some of the negative-degree terms in~\eqref{eq:Laurent}---the ``principal part'' of the Laurent expansion---in quantum gravity~\cite{Polyakov:1987zb,Knorr:2018kog,Borissova:2022clg}, black hole physics~\cite{Knorr:2022kqp}, and cosmology~\cite{Wetterich:1997bz, Deser:2007jk, Nersisyan:2016jta,Vardanyan:2017kal,Belgacem:2020pdz}, investigating their impact on black hole thermodynamics is of great importance. 

In order to compute the entropy corrections stemming from this type of infrared non-localities, while avoiding to compute the corresponding dressed field equations and solutions, we need to make some assumptions. First, we assume that the length scale at which quantum gravity effects become important is of the order of the Planck length, $l_{QG}\simeq l_{P}$, such that no strong non-perturbative effects  happen at the horizon scale for large black holes. Second, we limit ourselves to astrophysical black holes, for which $M\gg M_P$. This ensures that the event
horizon is located within a few Planck lengths from the Schwarzschild radius, i.e., that $r_H\simeq r_s$, with $r_s\equiv 2MG$.

The solutions to the quantum theory associated with the effective action~\eqref{action} ought to be found by solving the quantum field equations
\begin{equation}\label{eq:fieldeqs}
    \frac{\delta \Gamma_\mathrm{eff}}{\delta g_{\mu\nu}}=0\,.
\end{equation}
These solutions are generally different than those obtained from the vacuum Einstein equations (Ricci-flat metrics)\footnote{There are two exceptions, i.e., when the effective Lagrangian is local and $d=4$, and when $\F_3=0$ \cite{Myung:2017axf}, as in these cases the equations of motion are proportional to $R_{\mu\nu}$. We will not make this assumption in this work.} and are typically difficult to derive.

Even if the field equations~\eqref{eq:fieldeqs} are different than in GR, they are expected to admit spherically symmetric solutions of the type\footnote{Some regular black holes can indeed be seen as solutions to gravitational theories involving inverse powers of the d'Alembertian~\cite{Knorr:2022kqp}.}
\begin{align}\label{metric}
    ds^2=-f(r)dt^2+\frac{dr^2}{g(r)}+r^2d\Omega^2\,,
\end{align}
where $d\Omega^2$ is the line element on the $2$-sphere. Further, for large black holes the metric coefficients $f(r)$ and $g(r)$ can be written as
\begin{align}\label{metric-expansion}
    f(r)\simeq 1-\frac{r_s}{r}+\frac{A}{r^\alpha}, \quad g(r)\simeq1-\frac{r_s}{r}+\frac{B}{r^\beta}\,,
\end{align}
with $\alpha,\beta\geq 2$, i.e., the metric only differs from the Schwarzschild one by sub-leading terms. In particular, $g\simeq f$ for large or massive black holes.

This geometry has a Killing horizon $\mathcal{H}$, with bifurcation surface $\Sigma$, where the future and past horizons intersect. This is characterized by an antisymmetric bifurcation tensor $\epsilon_{\mu\nu}$, normalized such that $\epsilon_{\mu\nu}\epsilon^{\mu\nu}=-2$, and given by
\begin{equation}\label{eq:bifurcationtensor}
    \epsilon_{\mu\nu}=\sqrt{\frac{f(r)}{g(r)}}(\delta_\mu^t\delta_\nu^r-\delta_\mu^r\delta_\nu^t)\,.
\end{equation}
A spacelike surface spanning from $\Sigma$ to spatial infinity $i^0$ can be used to define a gravitational phase space consistent with general covariance\footnote{Notice that the Schwarzschild-like coordinates become singular at the bifurcation surface~$\Sigma$. Either by extending these coordinates or by changing to a chart where the coordinates are regular at the bifurcation surface, it can be shown that the phase space and the entropy formula retain their form~\cite{Kay:1988mu}.}. Wald~\cite{Wald:1993nt} showed that the first law of black hole thermodynamics follows, and the entropy acquires the interpretation of a Noether charge. For a static\footnote{The formula is valid for stationary black holes, and it does not necessarily satisfy a second law~\cite{Jacobson:1994qe} in the presence of, e.g., perturbations (see however~\cite{Iyer:1994ys} for an alternative dynamical definition). Here we limit ourselves to the static case.}, spherically-symmetric black hole it reads 
\begin{equation}\label{waldformula}
    S_W=-2\pi\int_{\Sigma}\left(\frac{\delta \mathcal{L}}{\delta R_{\mu\nu\rho\sigma}}\right)\epsilon_{\mu\nu}\epsilon_{\rho\sigma}dV_2^2\,,
\end{equation}
where $\mathcal{L}$ is the (classical or effective) Lagrangian, $dV_2^2=r^2\sin\theta d\theta d\phi$ is the line element on the bifurcation surface $\Sigma$, and the functional derivative is performed at a fixed metric. We shall employ this formula to compute the corrections to the entropy of the spherically symmetric spacetime~\eqref{metric-expansion} stemming from the quadratic higher-derivative corrections~\eqref{hd-action} with form factors structurally given by the principle part of the expansion~\eqref{eq:Laurent}. \newpage


\emph{Localizing the effective action} 
--- In the derivation of the Wald entropy formula, there are several assumptions that deserve to be highlighted: \textit{(i)} The theory is defined by a \textit{locally constructed}~\cite{Wald:1990}, diffeomorphism invariant Lagrangian $\mathcal{L}$ built from dynamical fields and their derivatives on a Lorentzian manifold. \textit{(ii)} There is some notion of asymptotic flatness, so that the black hole horizon can be defined as the (inner) boundary of the past region of the asymptotic region. \textit{(iii)} The horizon is a Killing horizon, i.e., a null surface to which a Killing field is normal.

Out of these assumptions, the last one is always satisfied for static black holes, and in particular for all spherically symmetric solutions. The property of asymptotic flatness is necessary to precisely define the notion of an event horizon, but given that such a construction is possible, it does not enter later in the derivation\footnote{The entropy formula at the horizon is still valid in non-asymptotically flat scenarios~\cite{Brustein:2009wr}. However, note that Wald's derivation of the first law of black hole thermodynamics depends on the asymptotic structure. This enters in the definition of asymptotic mass or angular momentum, for example.}. The most critical assumption for our purposes is the first one, since the action used is not directly ``locally constructed''. This is a problem because the covariant phase space formalism relies on the properties of jet bundles on field space~\cite{Wald:1990}. There is however a way to circumvent this issue: it consists in finding a ``localized'' version of the action that on-shell yields the same  configurations. Such a localized effective action is derived by translating the non-locality into a set of constrained auxiliary fields~\cite{Barvinsky:2011rk,Deser:2013uya,Pestun:2016jze,Teimouri:2017xqn}.

In order to proceed further, we now focus on the principal part of the expansion~\eqref{eq:Laurent}, and in particular on the form factors~$\F_i(\Box)$ given by the following finite sums
\begin{equation}\label{eq:formfactors}
    \F_i(\Box)=\sum_{n=1}^N F_{i,n}\,,\qquad F_{i,n} \equiv c_{i,n} \Box^{-n}\,,
\end{equation}
where $c_{i,n}$ are real coefficients and the truncation order~$N$ can be systematically increased. 
Following standard procedures~\cite{Barvinsky:2011rk,Deser:2013uya,Pestun:2016jze,Teimouri:2017xqn}, it is straightforward to compute a localized version of the effective action, $\Gamma^{\mathrm{local}}_{\mathrm{eff}}=\Gamma_\mathrm{EH}+ \Gamma_{\textrm{eff,HD}}^{\textrm{local}}$, made up of the Einstein-Hilbert part $\Gamma_\mathrm{EH}$ and the localized higher-derivative action (see supplemental material). We will use such an action to compute the entropy corrections stemming from the form factors~\eqref{eq:formfactors}.


\emph{Entropy formula} --- The entropy of large black holes in a theory with infrared non-localities~\eqref{eq:formfactors} is to be computed as a Noether charge, Eq.~\eqref{waldformula}, with the action $S\equiv\Gamma^{\mathrm{local}}_{\mathrm{eff}}=\Gamma_\mathrm{EH}+ \Gamma_{\textrm{eff,HD}}^{\textrm{local}}$ being the localized version of the effective action. After some algebra (see supplemental material for details), the resulting dimensionless entropy~$\tilde{S}_W=(4G/\mathcal{A})S_W$ takes the form
\begin{equation}\label{entropyformula}
\begin{aligned}
    \tilde{S}_W & =\lim_{r\to r_H} \biggl(  1+2\F_1(\Box) R_1 + \F_2(\Box)R_2-4\frac{g}{f}\F_3(\Box) R_3\\ 
    &  -\sum_{n=1}^N \left(\xi_{n, (1)} R_1-\xi_{n, (2)}R_2-4\frac{g}{f}\xi_{n,(3)}R_{3}\right)\biggr)\,.
\end{aligned}
\end{equation}
where $\xi_{n,(i)}$ are Lagrange multipliers, and 
\begin{equation}\label{eq:Curvature-Invariants-main}
\begin{aligned}
    R_1&=\frac{g f'^2}{2f^2}-2\frac{-1+g+rg'}{r^2}-\frac{rf'g'+2g(2f'+rf'')}{2rf}, \\
    R_2&=\frac{rg(f')^2-2f^2g'-f[rf'g'+2g(f'+rf'')]}{2rf^2}\,,\\
    R_{3}&=\frac{1}{4}\left(\frac{-f'^2}{f}+\frac{f'g'}{g}+2f''\right)\,.
\end{aligned}
\end{equation}
are functionals of the metric components~$f(r)$ and~$g(r)$. 

\emph{Diverging black hole entropy} --- Exploiting the generalized formula~\eqref{entropyformula}, we now compute the corrections to the Bekenstein-Hawking area law stemming from the form factors~\eqref{eq:formfactors}. As the entropy is additive, corrections stemming from the individual~$\Box^{-n}$-operators in the form factors~\eqref{eq:formfactors} can be determined individually. We checked that the corrections in the second line of Eq.~\eqref{entropyformula} are finite and unimportant to our conclusions. In the following we will thus focus on the dimensionless contributions 
\begin{equation}\label{eq:dimentropycontr}
\begin{aligned}
     \tilde{S}_W^{(n)} \equiv \lim_{r\to r_H}\left( 2 c_{1,n}\frac{1}{\Box^{n}}R_1+c_{2,n}\frac{1}{\Box^{n}}R_2-4c_{3,n}\frac{g}{f}\frac{1}{\Box^{n}}R_3 \right) \, .
\end{aligned}
\end{equation}
To this end, we first need to determine how the $\Box^{-n}$ terms act on the radial functions $R_i(r)$. This requires solving $\Box^{n}\psi(r)=R_i(r)$ with respect to the function $\psi$. For a generic spherically symmetric metric~\eqref{metric}~\cite{Knorr:2022kqp}
\begin{equation}\label{eq:oneoverboxformula}
\begin{aligned}
    {\Box^{-1}}\phi(r)=\Phi[\phi,r,R_x,R_y]\,,
\end{aligned}
\end{equation}
where $\phi$ is a generic function of the radial coordinate and we have defined
\begin{equation}\label{eq:defPhi}
\begin{aligned}
     \Phi[\phi,r,R_x,R_y]\equiv \int_r^{R_x} \hspace{-0.2cm}\int_x^{R_y} dx\, dy \frac{-y^2\phi(y)}{x^2\sqrt{g(x)f(x)}} \sqrt{\frac{f(y)}{g(y)}}\,.
\end{aligned}
\end{equation}
This is the general solution to Eq.~\eqref{eq:oneoverboxformula}, and the ``cutoffs'' $R_x$ and $R_y$ are related to its initial conditions. As in the case $n=1$ we have $\phi(r)=R_i(r)$, the regularity properties of $R_i(r)$ and the absence of zero modes allow selecting special initial conditions, such that $R_x,R_y\to\infty$~\cite{Knorr:2022kqp}. We will nevertheless show that our results are independent of the choice of initial conditions.
Eq.~\eqref{eq:defPhi} can be generalized to a recursion formula where $\phi_{n+1}\equiv\Box^{-n-1} \phi(r)$ is written in terms of~$\phi_n\equiv\Box^{-n} \phi(r)$,
\begin{equation}\label{eq:recursionformula}
    \phi_{n+1}(r)=\Phi[\phi_n,r,R_x^{(n)},R_y^{(n)}]\,,
\end{equation}
where $\phi_n\equiv \Box^{-1} \phi_{n-1}=\Box^{-n}\phi(r)$ with $n\geq1$, and $\{R_x^{(n)},R_y^{(n)}\}$ are initial conditions.
With this, the starting point to compute the corrections to the area law is to consider the lowest-order operators, $F_{i,1} = c_{i,1} / \Box$, and then compute the effect of the others---the $F_{i,n} = c_{i,n} \Box^{-n}$ in Eq.~\eqref{eq:formfactors}---recursively. We anticipate that the divergence of the lowest-order operator implies, owed to the recursion formula~\eqref{eq:recursionformula}, the divergence of all the higher-order terms. 

The first three contributions to the entropy formula that we have to compute are thus
\begin{equation}\label{eq:maincorrections}
\begin{aligned}
    &\mathcal{A}_{i,1}\equiv F_{i,1} R_i = \Phi[c_{i,1}R_i,r_H,R_x,R_y] \,,\quad i=1,2,3 \, ,
\end{aligned}
\end{equation}
where we have considered the same $(R_x,R_y)$ $\forall i = 1,2,3$ for simplicity. This is justified since, as we shall see, our conclusions are independent of the initial conditions.

Based on our assumptions, the solution to the non-local effective field equation~\eqref{eq:fieldeqs} is of the form~\eqref{metric-expansion}. Since $g\simeq f$ for large and massive black holes, in what follows we shall set $A=B$ and $\alpha=\beta$. We checked that this simplification does not affect our conclusions.

It is useful to start from a simple example, where exact analytical formulas can be derived: the Schwarzschild black hole, corresponding to the case $A=0$, is an exact solution for ${\mathcal{F}}_3=0$. In this ideal case $\mathcal{A}_{3,1}=0$ by construction, while the other two contributions in Eq.~\eqref{eq:maincorrections} can be computed with the aid of the formula~\eqref{eq:defPhi}. In particular, $\mathcal{A}_{1,1}=0$ and 
\begin{equation}
    \begin{aligned}
        &\Phi[R_2,r,R_x,R_y]  \\
        &= \frac{1}{2}\log\left(\frac{R_y}{R_x}\right)\log\left(\frac{R_y(r_s-R_x)^2}{R_x r_s^2}\right) -\mathrm{Li}_2\left(\frac{R_x}{r_s}\right)\\
        &-\frac{1}{2}\log\left(\frac{R_y}{r}\right)\log\left(\frac{R_y(r_s-r)^2}{rr_s^2}\right)+ \mathrm{Li}_2\left(\frac{r}{r_s}\right)\, ,
    \end{aligned}
\end{equation}
where $\mathrm{Li}_2$ is a polylogarithm of order two, which is real for $r\to r_H\leq r_s$. Analogous analytical formulas can also be derived in the case $\alpha=2$ (see supplemental material). Taking the limit $r\to r_H=r_s$, one finds that $\mathcal{A}_{2,1}$ diverges, with the sign of the divergence depending on the coefficient~$c_{2,1}$. In particular, divergences arise both at the horizon and at infinity (as $R_x,R_y\to\infty$).

\begin{figure}[t!]
    \includegraphics[width =0.9 \columnwidth]{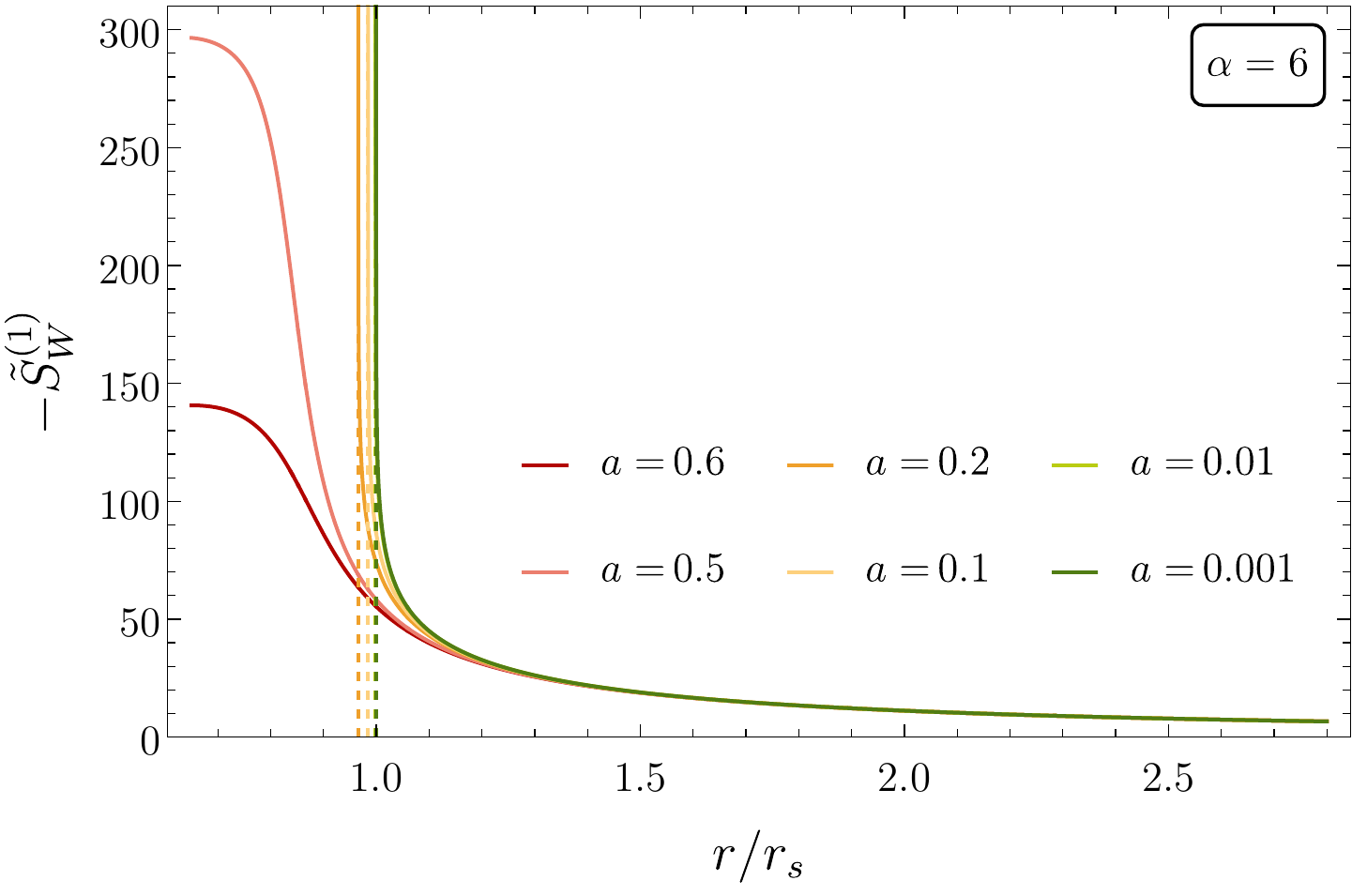}
    \caption{Dimensionless entropy correction $\Tilde{S}_W^{(1)}$ as a function of $r/r_s$, for $R_x = R_y = 10^3$, $c_{1,1}=c_{2,1}=c_{3,1}=1$, $\alpha=6$ and different values of~$a$. For each $a$, the location of the event horizon $r_H$ (when it exists) is displayed as a vertical dashed line. If an event horizon is present, the entropy diverges as~$r\to r_H$.}
    \label{fig:Divergence-rH-alpha6}
\end{figure}

\begin{figure}[t!]
\includegraphics[width = 0.9\columnwidth]{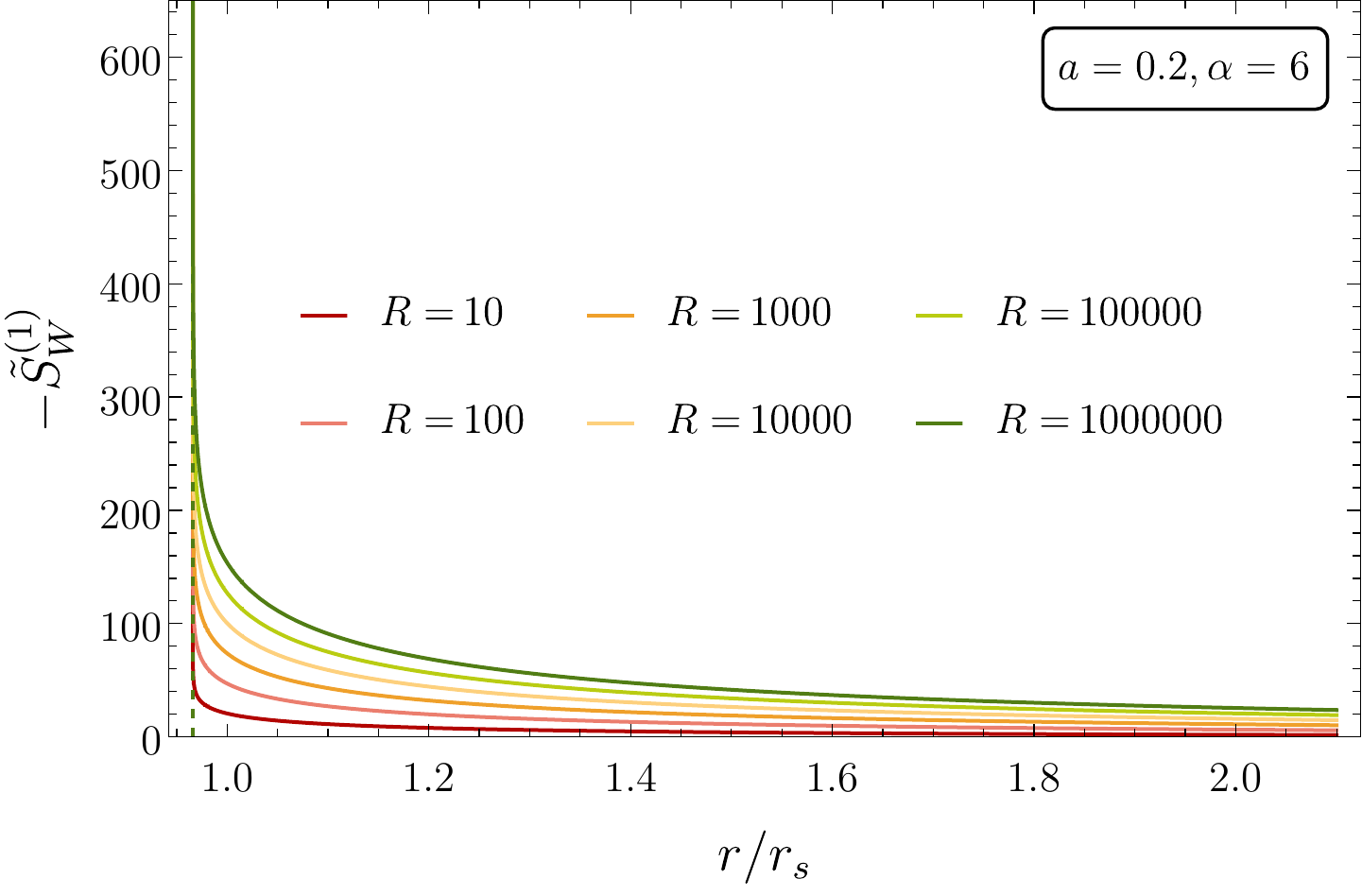} 
        \caption{Dimensionless entropy correction $\tilde{S}_W^{(1)}$, cf. Eq.~\eqref{eq:dimentropycontr}, for $a=0.2$, $\alpha=6$, $c_{1,1}=c_{2,1}=c_{3,1}=1$ and different values of $R=R_x=R_y$. The vertical dashed line denotes the position of the event horizon. The divergence of the entropy is independent of the initial conditions $R$. }
    \label{fig:Convergence-Rx}
\end{figure}

Similar findings also hold when accounting for the asymptotic corrections to the Schwarzschild metric coefficients, in which case $A,B\neq0$ in Eq.~\eqref{metric-expansion} and $r_H\neq r_s$. 
In these cases one has to account for all corrections~\eqref{eq:maincorrections}. The event horizon, when it exists, is located at
\begin{equation}
    r_H\simeq\left(1+\frac{a}{(a-1)\,\alpha-1}\right)r_s\,,
\end{equation}
where~$a$ is the dimensionless constant $a\equiv (\alpha+1) r_s^{-\alpha}A$, and $a\ll1$ for large black holes.
Independent of the exponent~$\alpha$ and of the value of $a$ in the allowed range, whenever a horizon exists, the contributions~\eqref{eq:maincorrections} to the entropy $\tilde{S}_W^{(1)}$ diverge as $r \to r_H$. This is shown in Fig.~\ref{fig:Divergence-rH-alpha6}, where we account for the position of the horizon for different values of $a$. The presence of a horizon for large black holes requires $a \ll 1$. Indeed, when $a$ is big enough, no horizon forms, and the integrals~\eqref{eq:maincorrections} are regular everywhere. Vice versa, when a horizon exists (dashed lines in Fig.~\ref{fig:Divergence-rH-alpha6}), the entropy correction diverges (see supplemental material for further analytical and numerical evidence). In particular, as shown in Fig.~\ref{fig:Convergence-Rx}, such a divergence is independent of the initial conditions~$(R_x,R_y)$. Finally, higher-order terms in Eq.~\eqref{eq:formfactors} lead to similar divergent contributions $\tilde{S}_W^{(n)}$ owed to the recursion formula~\eqref{eq:recursionformula}. 

Whether the divergence is positive or negative infinity depends on the
(relative) signs of the coefficients $c_{i,n}$, with $i = 1,2,3$ and $n=1,\dots,N$. As discussed in the supplemental material, even in the simple case $N=n=1$ and $\alpha=2$, the condition on the coefficients $c_{i,1}$ to eliminate the divergence is non-trivial.

Plugging all corrections $\tilde{S}_W^{(n)}$ into Eq.~\eqref{entropyformula}, and considering that the  terms in the second line of Eq.~\eqref{entropyformula} yield finite contributions, we finally obtain the surprising result that the total black hole entropy $S_W$ diverges,
\begin{equation}
    S_W \to \pm\infty\,,
\end{equation}
unless the set of coefficients $c_{i,n}$ are fine-tuned such that all divergences generated by all $\Box^{-n}$ operators in the form factors~\eqref{eq:formfactors} cancel out. Such a special combination of coefficients is unlikely realized by the quantum gravitational dynamics.

Identifying the Wald entropy with the thermodynamical entropy given by the Boltzmann formula\footnote{This identification does not hold if the black hole interior carries a non-zero entropy~\cite{Rovelli:2017mzl, Rovelli:2019tbl}. In this case the total, von Neumann entropy would be the sum of the Wald entropy and that associated with the internal degrees of freedom. The number of degrees of freedom would then be related to the von Neumann entropy.} 
\begin{equation}
    S_W \equiv S_B=k_B \log{W} \,,
\end{equation}
our result entails that spherically symmetric configurations stemming from effective actions that display infrared non-localities are either highly chaotic or thermodinamically impossible. In turn, our result places strong constraints on the Laurent expansion of the gravitational effective action: finiteness of the black hole entropy requires the absence of quantum-induced infrared non-localities, or a very special set of coefficients resulting from quantum gravitational dynamics.


\emph{Conclusions} --- In our Letter, we tackled the problem of determining corrections to the black hole area law stemming from infrared non-localities in the gravitational effective action. Higher-derivative corrections to the effective action are generally expected from quantum gravity and are particularly important in the physics of large black holes, as a description in terms of an effective action has to be recovered irrespective of the specific ultraviolet completion of gravity. Local higher-derivative terms have been considered in the seminal work~\cite{Conroy:2015wfa}, and yield sub-leading contributions to the black hole area law. Our Letter complements the results in~\cite{Conroy:2015wfa} by accounting for non-local higher derivative terms, the latter being of relevance in several quantum gravity models~\cite{Polyakov:1987zb,Knorr:2018kog,Borissova:2022clg}, black hole physics~\cite{Knorr:2022kqp}, and cosmology~\cite{Wetterich:1997bz, Deser:2007jk, Nersisyan:2016jta,Vardanyan:2017kal,Belgacem:2020pdz}. Despite some of these infrared non-localities seem compatible with observations~\cite{Nersisyan:2016jta,Vardanyan:2017kal,Belgacem:2020pdz}, they appear to yield inconsistencies: we find that not only infrared non-localities lead to corrections that are dominant over the Bekenstein-Hawking term, as we expected, but they also yield diverging contributions. Exploiting the Boltzmann formula, one would then conclude that the corresponding black holes are made of either zero or infinitely many microstates, depending on the sign of the divergence. This paradoxical behavior points to novel constraints on the Laurent expansion of the effective action stemming from quantum gravity theories: its principle part has to vanish or be fine-tuned in order for the black hole entropy to remain finite. 

It is still an open question whether similar cubic and higher-order terms in the curvature expansion~\eqref{action} of the effective action could dramatically change this conclusion, and whether resumming all infinitely many corrections could provide a finite result. Yet, this is expected to come at the expense of fine-tuning infinitely many coefficients in the effective action.


\begin{acknowledgments}
The authors would like to thank I. Basile, B. Knorr, D. Pere\~{n}iguez, and A. Riello for interesting discussions, and I. Basile, B. Knorr, and V. Cardoso for feedback on our manuscript. The authors acknowledge support by Perimeter Institute for Theoretical Physics. Research at Perimeter Institute is supported in part by the Government of Canada through the Department of Innovation, Science and Economic Development and by the Province of Ontario through the Ministry of Colleges and Universities. JRY acknowledges support from the Villum Investigator program supported by the VILLUM Foundation (grant no. VIL37766) and the DNRF Chair program (grant no. DNRF162) by the Danish National Research Foundation. AP also acknowledges Nordita for support within the ``Nordita Distinguished Visitors'' program and for hospitality during the last stages of development of this work. Nordita is supported in part by NordForsk.
\end{acknowledgments}

\bibliographystyle{apsrev4-2}
\bibliography{biblio.bib}

\clearpage


\renewcommand{\thesubsection}{{S.\arabic{subsection}}}
\setcounter{section}{0}

\section*{Supplemental material}

In this appendix we provide technical details on the derivation of the entropy formula that have been omitted in the main text, as well as further analytical and numerical evidence supporting our results.

\subsection{Localization of the effective action}

We start our derivation by detailing the localization procedure of the effective action. The localization entails introducing a number of auxiliary fields satisfying specific constraints that are enforced by a set of Lagrange multipliers~\cite{Barvinsky:2011rk,Deser:2013uya,Pestun:2016jze,Teimouri:2017xqn}. In our case, the localization requires introducing $N$ scalars $\psi_{n,(1)}$, $N$ two-tensors $\psi^{\mu\nu}_{n,(2)}$, and~$N$ four-tensors $\psi^{\mu\nu\rho\sigma}_{n,(3)}$. The localized action reads
\begin{equation}\label{local-action}
\begin{aligned}
        \Gamma_{\textrm{eff,HD}}^{\textrm{local}}&=\frac{\alpha}{16\pi G}\int d^4 x \sqrt{-g}\,\sum_{n=0}^N \Bigl(c_{n,1} R \psi_{n, (1)}\\
        & +R\, \xi_{n, (1)}(\Box^n \psi_{n, (1)} -R) +c_{n,2} R_{\mu\nu} \psi^{\mu\nu}_{n, (2)} \\
        &  +R_{\mu\nu}\xi_{n, (2)}(\Box^n \psi^{\mu\nu}_{n, (2)} -R^{\mu\nu})+c_{n,3} R_{\mu\nu\rho\sigma} \psi^{\mu\nu\rho\sigma}_{n, (3)}\\
        &  +R_{\mu\nu\rho\sigma}\xi_{n, (3)}(\Box^n \psi^{\mu\nu\rho\sigma}_{n, (3)} -R^{\mu\nu\rho\sigma}) \Bigr)\,.
\end{aligned}
\end{equation}
The $3N$ scalars $\xi_{n,(i)}$ are Lagrange multipliers introduced in such a way that they enforce the on-shell constraints
\begin{equation}\label{eq:EOM}
\begin{split}
    \Box^n \psi_{n, (1)}-R&=0\,,\\
    \Box^n \psi^{ab}_{n, (2)}-R^{ab}&=0\,,\\
    \Box^n \psi^{abcd}_{n, (3)}-R^{abcd}&=0\,.
\end{split}
\end{equation}
One can check that when enforcing these constraints, as well as the field equations for the metric field, the localized action reproduces the original non-local one.  

\subsection{Derivation entropy formula}

In this subsection we derive the expression~\eqref{entropyformula} starting from the Wald entropy formula~\eqref{waldformula}. For a general spherically-symmetric spacetime~\eqref{metric}, the bifurcation tensor is given in terms of the metric functions as in Eq.~\eqref{eq:bifurcationtensor}.

For $f(r)=g(r)$ one can prove the exact relation
\begin{equation}\label{eq:commutationRepsilon}
    (\Box R^{\mu\nu\sigma\rho})\epsilon_{\mu\nu}\epsilon_{\sigma\rho}\equiv \Box(R^{\mu\nu\sigma\rho}\epsilon_{\mu\nu}\epsilon_{\sigma\rho})\,,
\end{equation}
i.e., $[\Box,\epsilon_{\mu\nu}]=0$. One can then use the same relation to show that higher powers of the d'Alembertian also commute with the bifurcation tensor. The argument also goes through for inverse powers of the d'Alembertian, as the commutator $[A,B]$ of an operator $A$ with $B$ and its inverse are proportional,
\begin{equation}
    [B,A^{-1}]=-A^{-1}[B,A]A^{-1}\,.
\end{equation}
Since asymptotically $f\sim g$, Eq.~\eqref{eq:commutationRepsilon} can be exploited to substantially simplify computations.

On this basis, the variation of the EH and HD Lagrangians in Eq.~\eqref{action} with respect to the Riemann tensor at a fixed metric is obtained by following standard steps (see, e.g.,~\cite{Teimouri:2017xqn}). We illustrate this procedure by explicitly showing the calculation of the Ricci term
\begin{equation}
    \begin{split}
        \frac{\partial \mathcal{L}_{\textrm{Ricci}}^{\textrm{Local}}}{\partial R_{\mu\nu\rho\sigma}}\epsilon_{\mu\nu}\epsilon_{\rho\sigma}&=\sum_{n=1}^N \left(c_{2, n} \frac{\delta R_{ab}}{\delta R_{\mu\nu\rho\sigma}}\epsilon_{\mu\nu}\epsilon_{\rho\sigma}\psi_{n, (2)}^{ab}\right.\\
        &\left.+\xi_{n, (2)}\frac{\delta R_{ab}}{\delta R_{\mu\nu\rho\sigma}}\epsilon_{\mu\nu}\epsilon_{\rho\sigma} (\Box^n\psi_{n, (2)}^{ab}-2R^{ab})\right)\\
        &=-\frac{1}{2}\left(\F_2(\Box)R_2-\sum_{n=1}^N \xi_{n, (2)}R_2\right)\,,
    \end{split}
\end{equation}
where the curvature invariant relevant for the entropy is
\begin{equation}\label{eq:R2_invariant}
    \begin{split}
    R_2&=\frac{rg(f')^2-2f^2g'-f\Bigl(rf'g'+2g(f'+rf'')\Bigr)}{2rf^2}\,.
    \end{split}
\end{equation}
The localization of the action modifies the entropy by a contribution that is proportional to the $\xi_{n,(i)}$ factors, but involving the same curvature invariant. We can anticipate from this expression that our result for the diverging black hole entropy does not depend on the localization procedure since, as we will show later, the first terms~$\F_i(\Box)R_i$ source the divergence whereas the term depending on the $\xi_{n,(i)}$ is finite. This contrasts with the procedure followed in~\cite{Teimouri:2017xqn}, where the Lagrange multipliers are fixed to be~$\xi_{n, (i)}=-c_{i,n}$; despite allowing for some simplifications, this choice could obscure the dependence of the result on the localization procedure. We shall thereby leave them unspecified.

Repeating this procedure for all terms in the Lagrangian and integrating over the bifurcation surface we obtain that the entropy is given by $S_W=(\mathcal{A}/4G)\tilde{S}_W$, where $\mathcal{A}=4\pi r_H$ is the black hole area,  $r_H$ stands for the event horizon, and the dimensionless entropy $\tilde{S}_W$ reads
\begin{equation}
\begin{aligned}
    \tilde{S}_W & =\lim_{r\to r_H} \biggl(  1+2\F_1(\Box) R_1 + \F_2(\Box)R_2-4\frac{g}{f}\F_3(\Box) R_3\\ 
    &  -\sum_{n=1}^N \left(\xi_{n, (1)} R_1-\xi_{n, (2)}R_2-4\frac{g}{f}\xi_{n,(3)}R_{3}\right)\biggr)\,.
\end{aligned}
\end{equation}
The curvature invariants involved in the entropy formula can be computed directly for the metric~\eqref{metric} and read 
\begin{equation}\label{eq:Curvature-Invariants}
\begin{aligned}
    R_1&=\frac{g f'^2}{2f^2}-2\frac{-1+g+rg'}{r^2}-\frac{rf'g'+2g(2f'+rf'')}{2rf}, \\
    R_{3}&=\frac{1}{4}\left(\frac{-f'^2}{f}+\frac{f'g'}{g}+2f''\right)\,.
\end{aligned}
\end{equation}
By requiring the metric to be Schwarzschild-like, i.e, setting $f=g$, the invariants of~\cite{Myung:2017axf} are recovered. 

\subsection{Recursion formula for non-local corrections}

In this subsection we detail the procedure to determine the action of a $\Box^{-n}$ operator on a generic radial function. We first notice that the d'Alembertian operator acting on purely radial functions is
\begin{align}
    \Box=-\frac{1}{r^2}\sqrt{\frac{g(r)}{f(r)}}\partial_r\left(r^2\sqrt{g(r)f(r)}\partial_r\right)\,.
\end{align}
Given a function $\phi(r)$, computing $\phi_1(r)=\Box^{-1}\phi(r)$ is equivalent to solving the differential equation $\Box \psi(r)=\phi(r)$. Its general solution reads~\cite{Knorr:2022kqp}
\begin{align}\label{1overbox}
    \frac{1}{\Box}\phi(r)=\int_r^{R_x}dx \, \frac{-1}{x^2 \sqrt{g(x) f(x)}} \int_x^{R_y} dy \, \sqrt{\frac{f(y)}{g(y)}} y^2 \phi(y)\,,
\end{align}
where $(R_x,R_y)$ are related to the initial conditions of the above differential equation. If the function $\phi(r)$ is sufficiently regular at infinity, the absence of zero modes selects special initial conditions such that $R_x,R_y\to\infty$. This is the case when $\phi(r)=R_i(r)$, but not for higher orders, i.e., when $\phi=\Box^{-n}R_i$ with $n\geq 1$.
Setting $\phi_n=\Box^{-n}\phi$, one can  exploit the formula~\eqref{1overbox} to obtain a recursive relation to compute every $\phi_n$. It reads
\begin{align}
    \phi_{n+1} = \int_r^{R_y^{(n)}} dx \, \frac{-1}{x^2 \sqrt{g(x) f(x)}} \int_x^{R_x^{(n)}} dy \, \sqrt{\frac{f(y)}{g(y)}} y^2\phi_n(y)\,,
\end{align}
where $(R_x^{(n)},R_y^{(n)})$ specify the initial conditions at the step $n+1$. Thus, in order to understand the divergent behavior of the $\Box^{-n}$ correction, it is necessary to characterize the divergences of the $\Box^{-1}$ term. 

\subsection{Constraints on the horizon}

We have introduced a general spherically symmetric metric~\eqref{metric}, where the metric functions $f(r)$ and $g(r)$ are required to be approximately of the Schwarzschild form, according to the asymptotic expansion~\eqref{metric-expansion}. Since the focus of our work is large black holes, we can place constraints on the parameters of such an expansion.

We can safely assume the outermost horizon $r_{H}$ to be located in the proximity of the Schwarzschild radius, so that $r_H=r_s(1+\epsilon)$, with $|\epsilon|\ll1$.
First, to identify~$\epsilon$, we can use the condition that for static black holes~$f(r_{H})=0$, which to leading order in $\epsilon$ reads
\begin{equation}
    f(r_H)\simeq A r_s^{-\alpha} + (1 - A \alpha r_s^{-\alpha}) \, \epsilon \,.
\end{equation}
Thus, the spacetime can have an event horizon at
\begin{equation}\label{eq:horizon}
    r_H=r_s(1+\epsilon)\,,\qquad 
    \epsilon\simeq \frac{A}{A \alpha -r_s^{\alpha}}\,.
\end{equation}
If it exists, the quantum-corrected horizon is located inside the Schwarzschild radius, so that $\epsilon\leq0$. This implies that $A\in\left[0,{r_s^{\alpha}}/{\alpha}\right]$. 
Further, the condition that the horizon is close to the classical Schwarzschild radius $r_s$ is $|\epsilon|\ll1$ and entails $|A|\ll {r_s^{\alpha}}/{(\alpha+1)}$.
In terms of the dimensionless constant $ a\equiv (\alpha+1) r_s^{-\alpha} A$ 
the two constraints read
\begin{equation}\label{eq:conditions}
    a\ll1 \,,\qquad a\in\left[0,\frac{\alpha+1}{\alpha}\right]\,.
\end{equation}
While Eq.~\eqref{eq:horizon} and the conditions~\eqref{eq:conditions} allow to estimate the location of the event horizon of large black holes, its existence depends on the parameters $(a,\alpha)$ and has to be checked separately. The existence of a horizon is crucially related to the divergence of the entropy in the presence of infrared non-localities.

\subsection{Analytic formulas for $\alpha=2$}\label{sec:constraints}

In this subsection, we show analytically the divergent behavior of the corrections to the entropy for the simplest geometry that deviates from the Schwarzschild metric. We consider the spacetime described by the ansatz~\eqref{metric-expansion} with $A=B$ and $\alpha=\beta=2$. We obtain that the leading-order contributions to the entropy are given by
\begin{equation}
    \begin{aligned}
        &\Phi[R_1,r,R_x,R_y] = 0, \\
        &\Phi[R_2,r,R_x,R_y] = \frac{r_++r_-}{r_--r_+}\left( \log (R_x) \log \biggl(\frac{r_- \left(r_+-R_x\right)}{r_+ \left(r_--R_x\right)}\biggr)
        \right. \\
        &\left. +\log (r) \log \left(\frac{\left(r-r_-\right) r_+}{r_- \left(r-r_+\right)}\right) -\text{Li}_2\left(\frac{R_x}{r_-}\right)+\text{Li}_2\left(\frac{R_x}{r_+}\right)\right. \\
        &\left.
        +\log \left(R_y^2\right)\tanh ^{-1}\Bigl(\frac{r_-+r_+-2 R_x}{r_--r_+}\Bigr)+\text{Li}_2\left(\frac{r}{r_-}\right)\right.\\
        &\left.-\log \left(R_y^2\right)\tanh ^{-1}\Bigl(\frac{r_-+r_+-2 r}{r_--r_+}\Bigr)-\text{Li}_2\left(\frac{r}{r_+}\right)\right), \\
        &\Phi[R_3,r,R_x,R_y] = \frac{3}{2} \log \left(\frac{r^2 \left(A-\left(r_-+r_+\right) R_x+R_x^2\right)}{\left(A+r^2-\left(r_-+r_+\right) r\right) R_x^2}\right)\\
        &+\frac{1}{R_y(r_+-r_-)} \, \biggl(\left(6 A-\left(r_-+r_+\right) R_y(\log \left(R_y^2\right)-3)\right)\\
        & \times \Bigl(\tanh ^{-1}\Bigl(\frac{r_-+r_+-2 r}{r_--r_+}\Bigr)-\tanh ^{-1}\Bigl(\frac{r_-+r_+-2 R_x}{r_--r_+}\Bigr)\Bigr) \\
        &+\left(r_-+r_+\right) R_y \left( +\text{Li}_2\left(\frac{R_x}{r_+}\right)-\text{Li}_2\left(\frac{R_x}{r_-}\right)\right.\\
        &\left.-\log \left(R_x\right) \log \left(\frac{A-r_+ R_x}{A-r_- R_x}\right)+\text{Li}_2\left(\frac{r}{r_-}\right)\right.\\
        &\left. +\log (r) \log \left(\frac{A-r r_+}{A-r r_-}\right)-\text{Li}_2\left(\frac{r}{r_+}\right) \right)\biggr).
    \end{aligned}
\end{equation}
where $r_{\pm}=r_s/2\pm1/2\sqrt{r_s^2-4A}$ denotes the location of the outer and inner horizons, respectively, and $\mathrm{Li}_2$ is a polylogarithm of order two. We have evaluated everything at a generic $r$, to show how the divergences occur when evaluating $r \to  r_H=r_+$. Taking the limit 
\begin{equation}
\begin{aligned}
    \lim_{r \to r_H} \Phi[R_2, r, R_x, R_y] &= \infty, \\
    \lim_{r \to r_H} \Phi[R_3, r, R_x, R_y] &= -\infty.
\end{aligned}
\end{equation}
The divergent part of the entropy is therefore controlled by the contribution
\begin{equation}\label{DeltaS}
\begin{aligned}
    \Delta S_W \equiv \tilde{S}_W^{(1)}= \lim_{r \to r_H}  ( & c_{2,1}\Phi[R_2, r, R_x, R_y]\\
    - &  4c_{3,1}\Phi[R_3, r, R_x, R_y])\,.
\end{aligned}
\end{equation}
The sign of the divergence thereby depends on the particular combination of the coefficients~$c_{i,1}$. In particular, the divergence may vanish for some special combinations of~$c_{i,1}$. We define the critical ratio $c_{\rm crit}$ as the ratio~$c_{2,1}/c_{3,1}$ such that~$\Delta S_W = 0$. 

\begin{figure}[t!]
    \vspace{20pt}
    \includegraphics[width = 0.95\columnwidth]{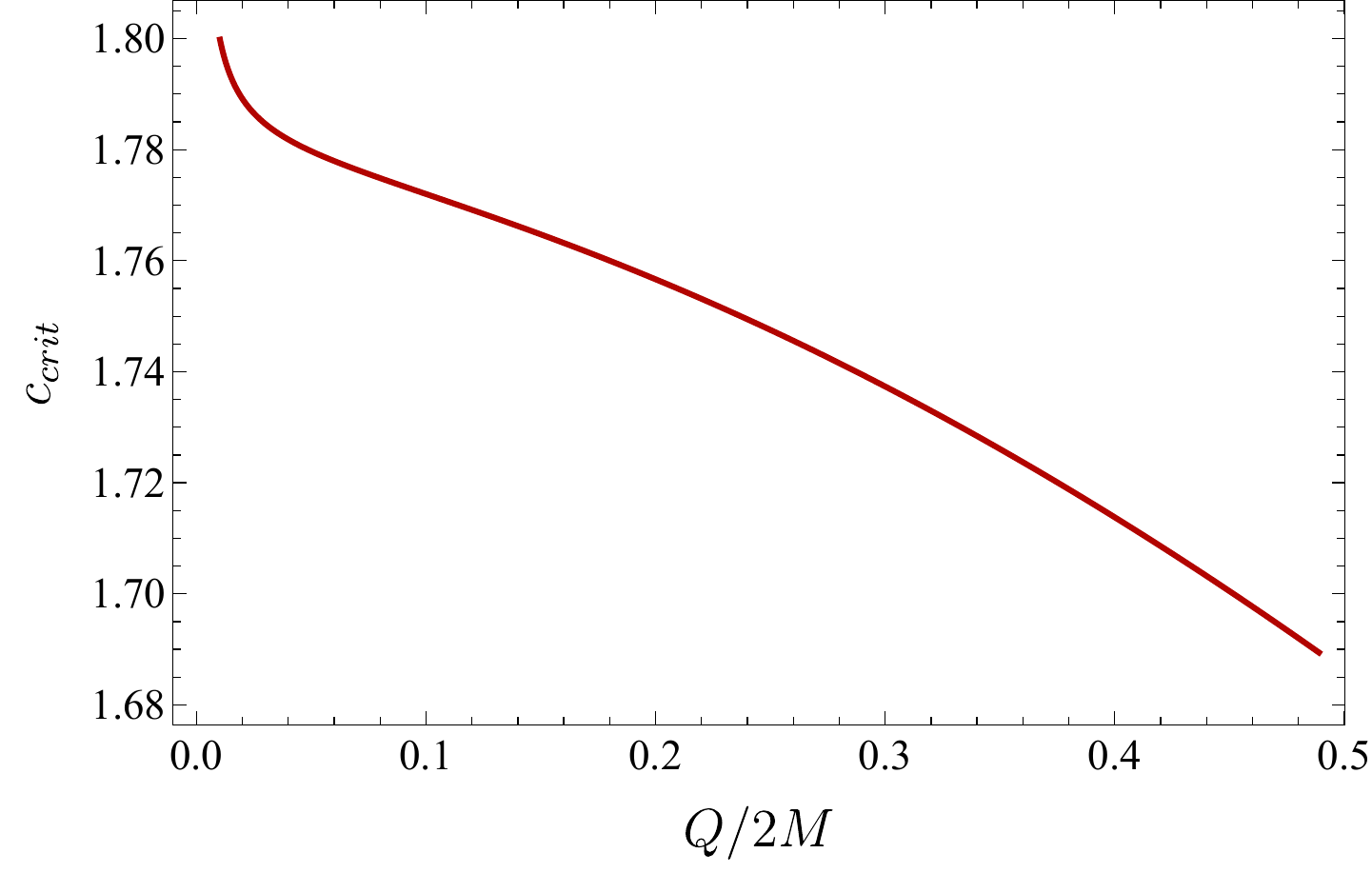}
    \caption{Critical value of the ratio between the coefficients $c_{2,1}$ and $c_{3,1}$ for which the entropy correction $\Delta S_W$ (c.f. Eq.~\eqref{DeltaS}) vanishes, as a function of the dimensionless charge $Q/2M$ of a Reissner-Nordstr\"om black hole of mass $M$. In the plot we fixed $R_x=R_y=10^3$. \label{fig:RN}}
\end{figure}

As an illustrative example, we study the critical ratio~$c_{crit}$ in the case of a Reissner-Nordstr\"om black hole. The metric for the solution to the classical Einstein-Maxwell theory, with charge~$Q$, is given by the ansatz~\eqref{metric-expansion} with $A = B =  Q$ and $\alpha = \beta = 2$. As for the Schwarzschild case, the corrections stemming from higher-order terms in the action could appear as sub-leading terms in the asymptotic expansion of the metric coefficients. As a first approximation, we shall neglect these sub-leading corrections. 
As shown by the above formulas, the entropy for Reissner-Nordstr\"om black holes in the presence of infrared non-localities diverges. In particular, we find that
\begin{equation}
    \Delta S_W = \begin{cases}
    \mathrm{sign}(c_{2,1})\times  \infty\,, \quad & c_{2,1} > c_{\mathrm{crit}}(Q/2M)\, c_{3,1}\,, \\
    -\mathrm{sign}(c_{2,1})\times \infty\,, \quad & c_{2,1} < c_{\mathrm{crit}}(Q/2M)\, c_{3,1}\,, \\
    \end{cases}
\end{equation}
where the critical threshold of the coefficients is a non-trivial function that is controlled by the charge of the black hole and its mass, as depicted in Fig.~\ref{fig:RN}.

\subsection{Parametric characterization of the divergence}

\begin{figure*}
\includegraphics[width = 0.97 \columnwidth]{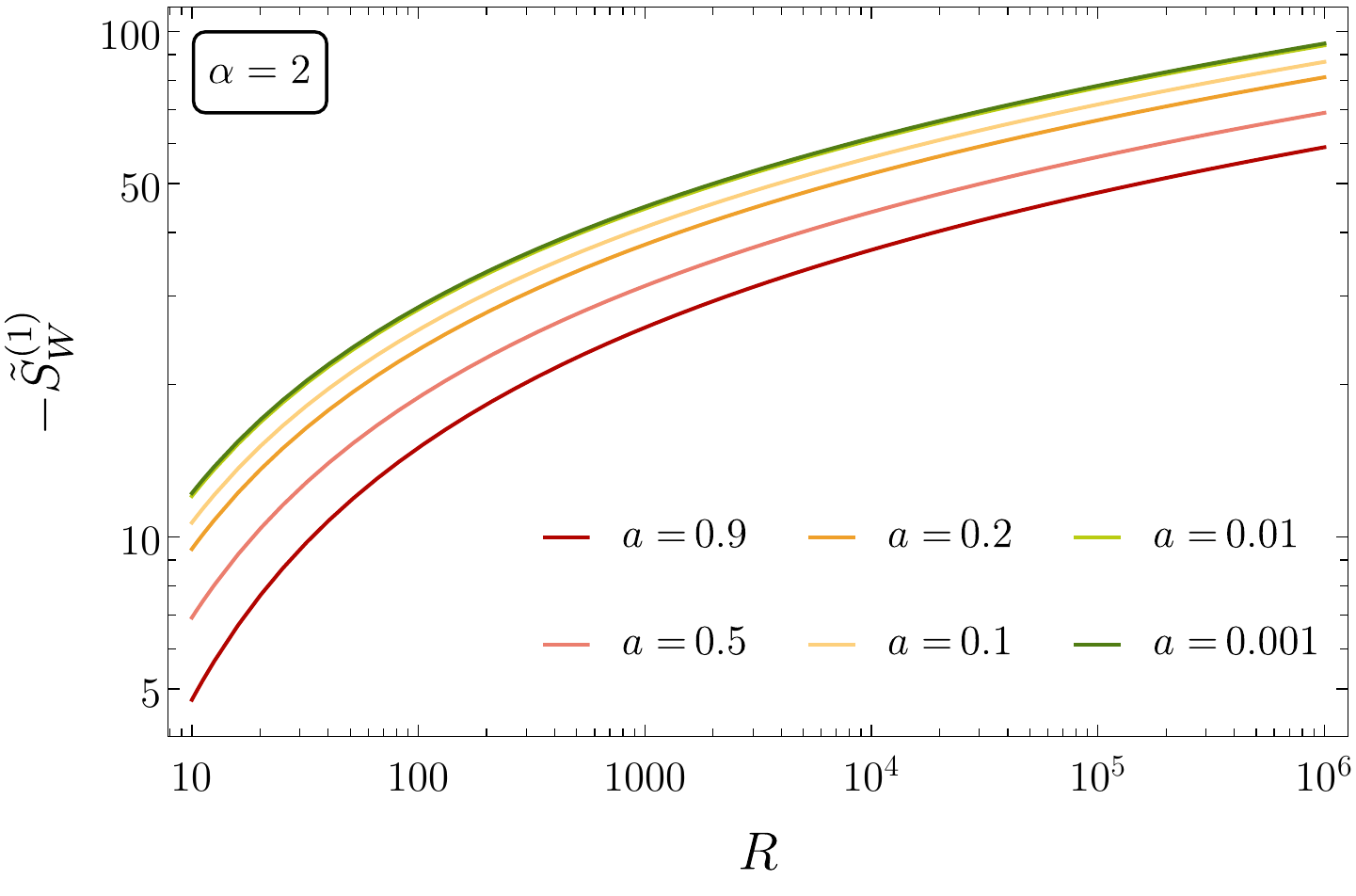} \hfill \includegraphics[width = \columnwidth]{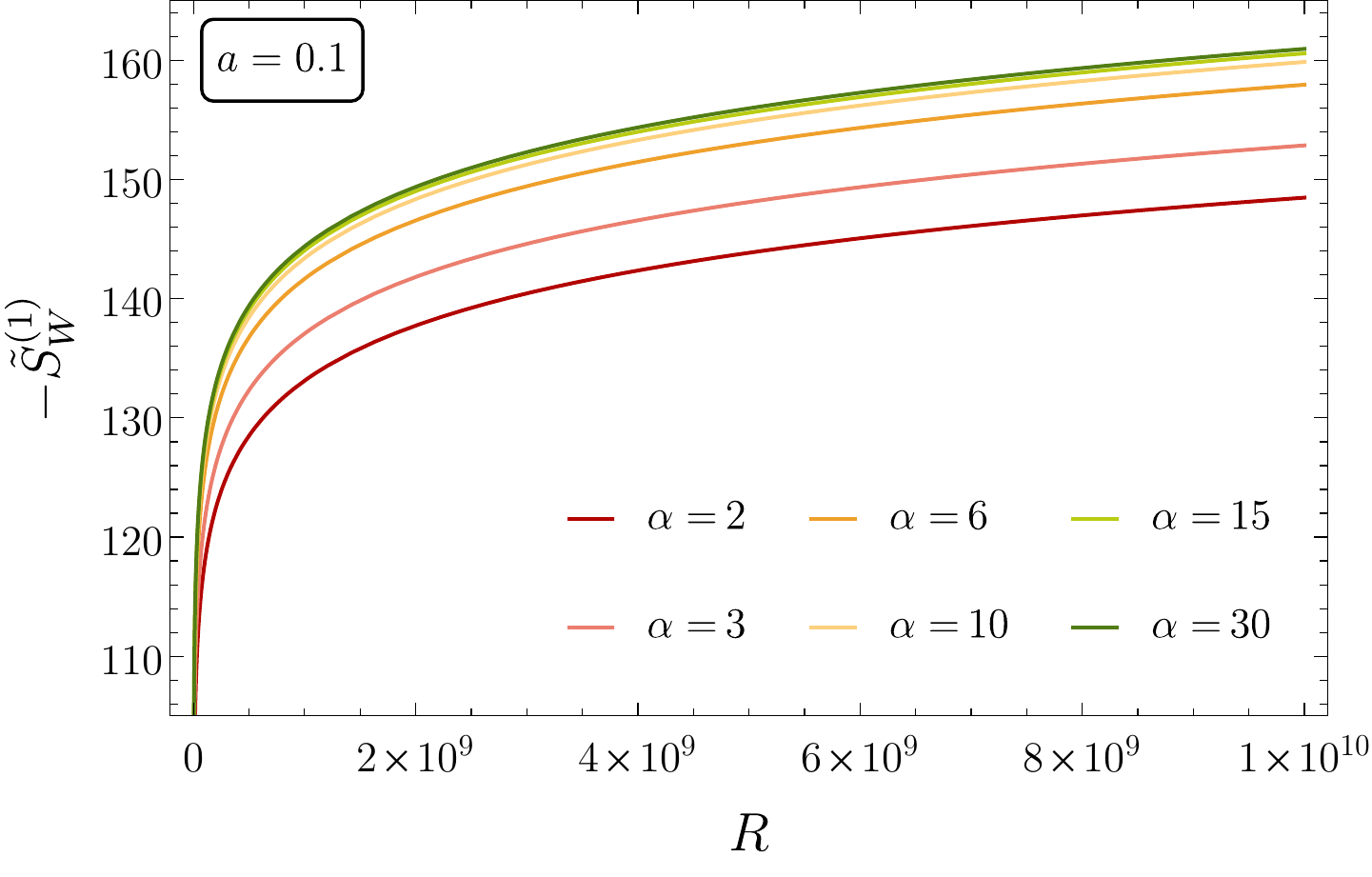} \caption{Dimensionless entropy correction $\tilde{S}_W^{(1)}$, cf. Eq.~\eqref{eq:dimentropycontr} as a function of the initial condition $R=R_x=R_y$. The correction is evaluated at $r =1.1\,r_s\neq r_H$ and for $c_{1,1}=c_{2,1}=c_{3,1}=1$. The left panel shows~$\tilde{S}_W^{(1)}$ for~$\alpha=2$ and different values of $a$, whereas the right panel depicts the same entropy correction for~$a=0.1$ and different values of the exponent~$\alpha$. In all these cases, the entropy diverges as $R\to \infty$. } \label{fig:Convergence-A}
\end{figure*}

We now extend the results from the previous subsection by numerically evaluating the  dimensionless entropy correction $\tilde{S}_W^{(1)}$, cf. Eq.~\eqref{eq:dimentropycontr}, in a large region of the parameter space spanned by $(a,\alpha)$.
\begin{figure*}[t!]
    \includegraphics[width = \columnwidth]{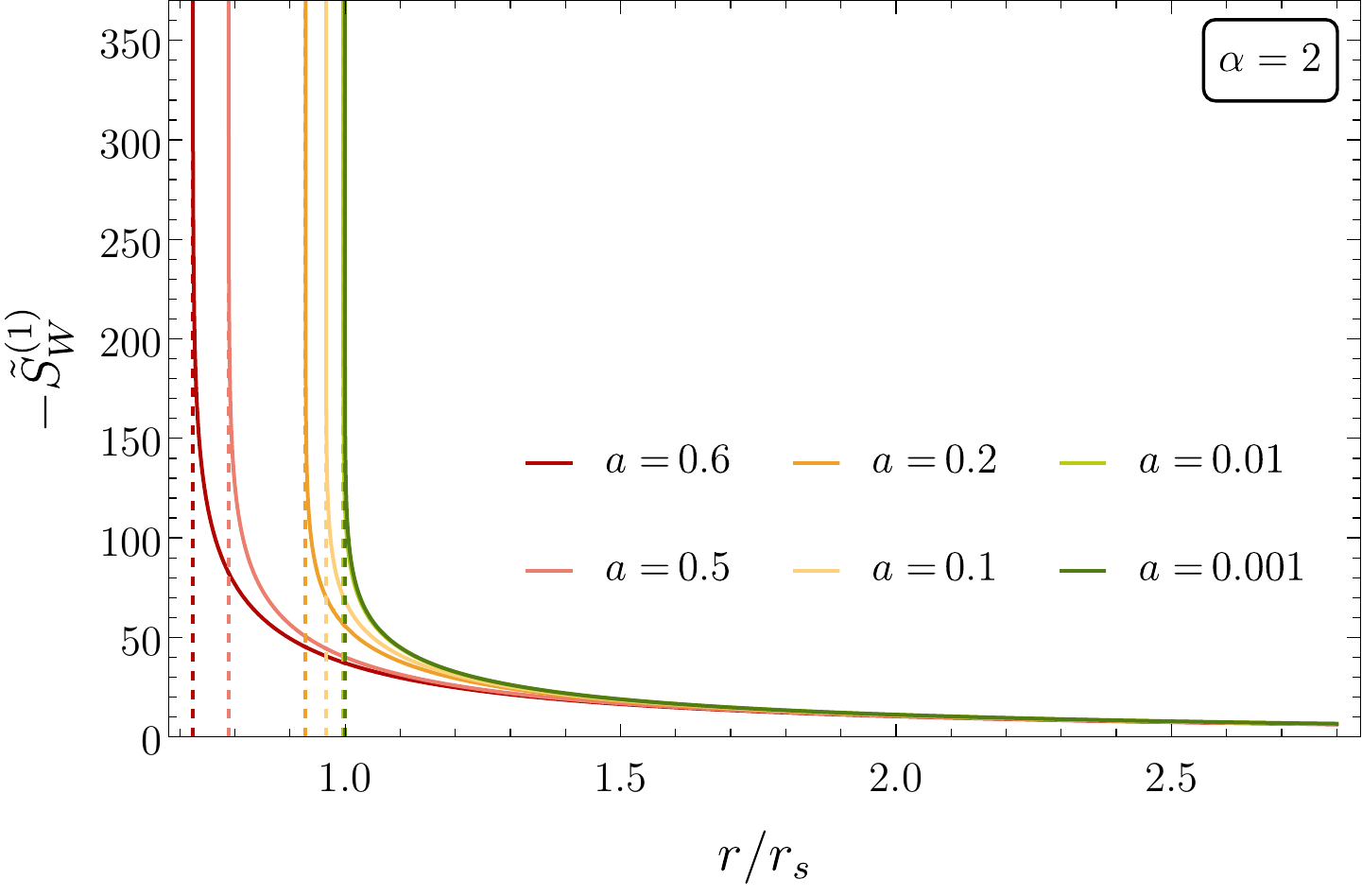}
    \includegraphics[width = \columnwidth]{Figures/Divergence_rS_n6.pdf}
    \includegraphics[width = \columnwidth]{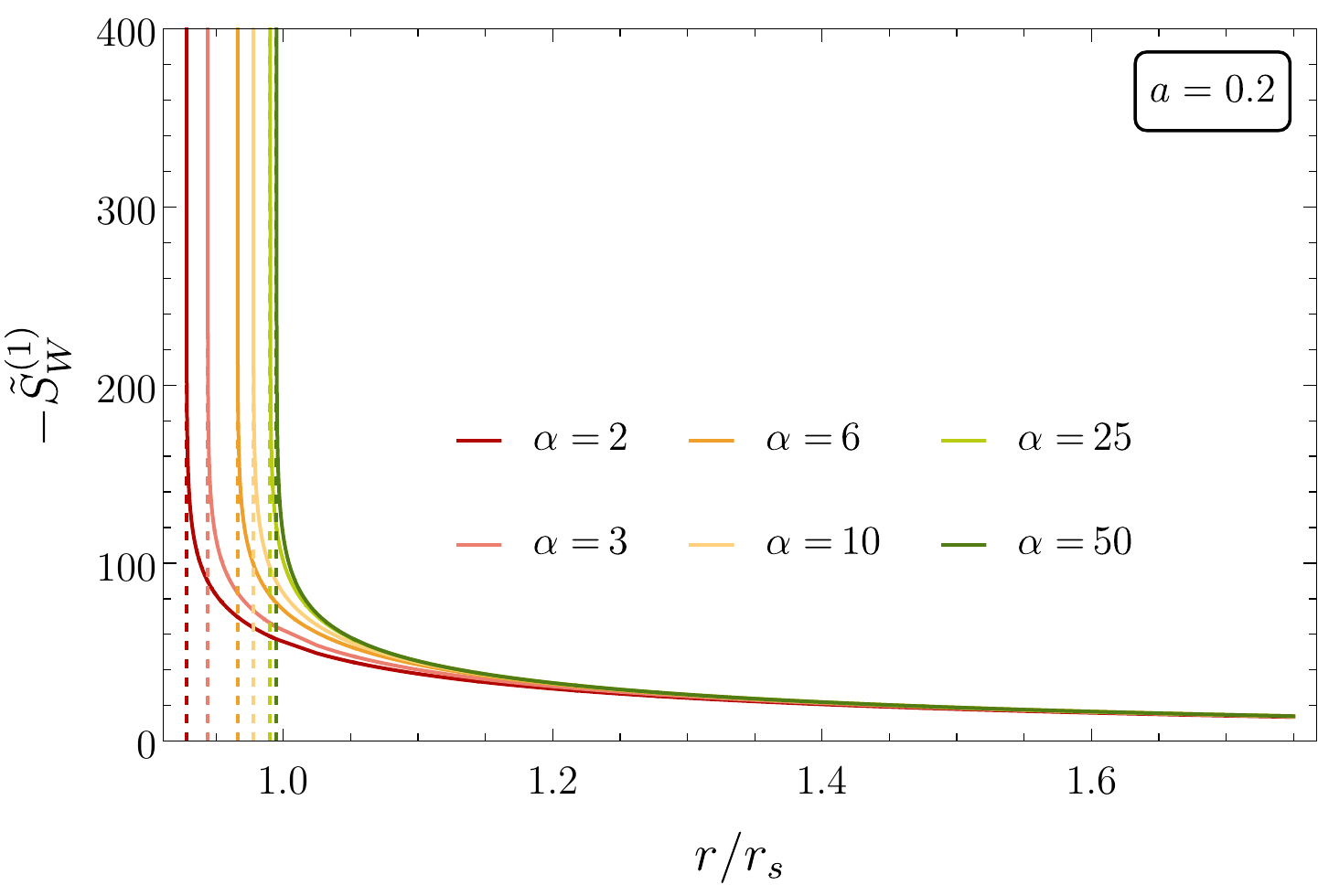}
    \includegraphics[width = \columnwidth]{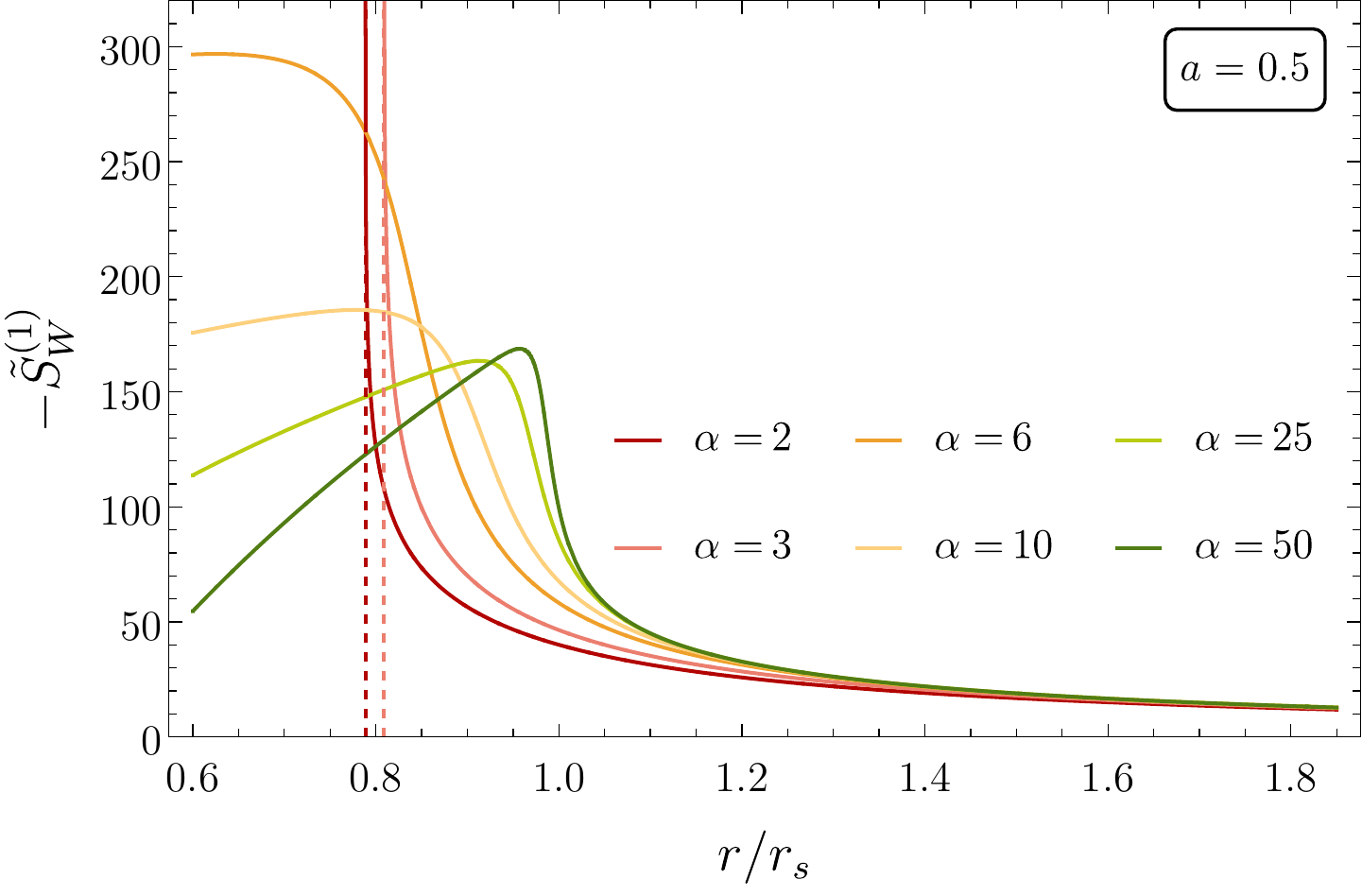}
    \caption{Dimensionless entropy correction $\Tilde{S}_W^{(1)}$ as a function of $r/r_s$, for $R_x = R_y = 10^3$, $c_{1,1}=c_{2,1}=c_{3,1}=1$, and various combinations of~$(a,\alpha)$, as reported in the figures. The location of the corresponding event horizons is indicated by a vertical dashed line. Provided that a horizon exists, the entropy diverges as~$r\to r_H$.} \label{fig:Divergence-alpha}
\end{figure*}
Some of the results were already reported in the main text, in Fig.~\ref{fig:Divergence-rH-alpha6} and~Fig.~\ref{fig:Convergence-A}. 

First, in Fig.~\ref{fig:Divergence-rH-alpha6} we fixed the value of the boundary condition~$R=R_x=R_y$ and showed that the corrected entropy diverges  as $r \to r_H$, provided that a horizon exists. Moreover, these features are independent of the exponent $\alpha$ and of the dimensionless coefficient $a$: the divergence as $r\to r_H$ is present whenever a horizon exists (see Fig.~\ref{fig:Divergence-alpha}). 

In addition, such a divergence is independent of the initial conditions characterized by $R = R_x = R_y$, as is clearly visible in Fig.~\ref{fig:Convergence-Rx}. We have also explicitly checked that the result also holds for $\alpha\neq\beta$ and $A\neq B$.

Secondly, in Fig.~\ref{fig:Convergence-A} we show that the corrected entropy with coefficients $c_{1,1}=c_{2,1}=c_{3,1}=1$ diverges to $\Tilde{S}_W^{(1)}\to -\infty$ as $R\to\infty$. Specifically, the divergence is faster for smaller values of $a$. More importantly, the divergent behavior does not depend on $a$, as long as it satisfies the constraints described previously, and is also independent of $\alpha$. 

We conclude from our analytical and numerical results that the divergent behavior, both as $r \to r_H$ and as $R \to\infty$ is independent of the exact values of $(a, \alpha)$, as long as the constraints on the horizon are satisfied, and modulo a special cancellation of divergences---as illustrated in the previous subsection. We emphasize that these results only account for the $\Box^{-1}$ correction in the form factors. If one were to include the whole series, there would be more divergent terms; hence a complicated relation between the $c_{i, n}$ coefficients would have to exist in order to potentially cancel all divergences. Such a cancellation is unlikely, as it would require a very special combination of Wilson coefficients in the gravitational effective action.

\end{document}